**Giant optical nonlinearity of Fermi polarons in atomically thin semiconductors**

Liuxin Gu[1], Lifu Zhang[1], Ruihao Ni[1], Ming Xie[2], Dominik S. Wild[3], Suji Park[4], Houk Jang[4], Takashi Taniguchi[5], Kenji Watanabe[5], Mohammad Hafezi[6], You Zhou[1,7,†]

[1]Department of Materials Science and Engineering, University of Maryland, College Park, MD 20742, USA

[2]Condensed Matter Theory Center, University of Maryland, College Park, MD 20742, USA

[3]Max Plank Institute of Quantum Optics, Garching, 85748, Germany

[4]Center for Functional Nanomaterials, Brookhaven National Laboratory, Upton, NY 11973, USA

[5]National Institute for Materials Science, 1-1 Namiki, Tsukuba 305-0044, Japan

[6]Joint Quantum Institute (JQI), University of Maryland, College Park, MD 20742, USA

[7]Maryland Quantum Materials Center, College Park, Maryland 20742, USA

†To whom correspondence should be addressed: youzhou@umd.edu

**Realizing strong nonlinear optical responses is a long-standing goal of both fundamental and technological importance[1-7]. Recently significant efforts have focused on exploring excitons in solids as a pathway to achieving nonlinearities even down to few-photon levels[8–11]. However, a crucial tradeoff arises as strong light-matter interactions require large oscillator strength and short radiative lifetime of the excitons, which limits their interaction strength and nonlinearity. Here we experimentally demonstrate strong nonlinear optical responses by exploiting the coupling between excitons and carriers in an atomically thin semiconductor of trilayer tungsten diselenide. By controlling the electric field and electrostatic doping of the trilayer, we observe the hybridization between intralayer and interlayer excitons along with the formation of Fermi polarons due to the interactions between excitons and free carriers[12]. We find substantial optical nonlinearity can be achieved under both continuous wave and pulsed laser excitation, where the resonance of the hole-doped Fermi polaron blueshifts by as much as ~10 meV. Intriguingly, we observe a remarkable asymmetry in the optical nonlinearity between electron and hole doping, which is tunable by the applied electric field. We attribute these features to the strong interactions between excitons and free charges with optically induced valley polarization. Our results establish that atomically thin heterostructures are a highly versatile platform for engineering nonlinear optical response with applications to classical and quantum optoelectronics[13–16], and open avenues for exploring many-body physics in hybrid Fermionic-Bosonic systems[17–19].**

Nonlinear optical phenomena lie at the heart of classical and quantum optics, with applications ranging from data storage and communications to quantum control[1-4]. Developing physical systems with stronger optical nonlinearity while reducing their power requirement holds the promise of more efficient optoelectronics and may unlock new technologies such as single-photon switches and transistors[4-7]. In recent years, significant efforts have been devoted to investigating excitons in semiconductors as a solid-state medium for realizing strong optical nonlinearity[8–11].

Van der Waals heterostructures based on atomically thin transition metal dichalcogenides (TMDs) have emerged as a new materials platform for fundamental studies of excitons and for engineering optical responses[20–23]. Excitons in such two-dimensional materials are highly tunable with rich spin-valley physics and possess characteristics promising for optical nonlinearity, such as strong light-matter interactions and weak screening of Coulomb potential[23,24]. However, a major challenge in achieving high nonlinearity under low excitation power arises from the balance between the strengths of exciton-photon and exciton-exciton interactions[16,25,26]. For instance, intralayer excitons in these materials exhibit large oscillator strength but experience weak interactions, dominated by exchange interactions and limited by their short lifetime[21]. On the other hand, interlayer excitons in TMD heterostructures have longer lifetimes and experience considerably stronger interactions due to their finite electric dipole moment[27,28]. Unfortunately, spatial separation of the electron-hole wavefunctions leads to weaker absorption of incident photons. Various strategies have been proposed to overcome these challenges, including generating polaritons to improve the intralayer exciton lifetime and hybridizing intra- and inter-layer excitons to enhance the interlayer exciton absorption[29-32]. In our study, we report giant nonlinear optical responses of Fermi polarons on the order of several millielectronvolts under photon flux of $10^{13}$ photons per second per square micrometer, which are highly tunable by doping and electric field. We attribute the tunable nonlinear optical response to the strong interactions between bright intralayer excitons and free carriers, which introduces a net valley polarization in the materials, which forms a novel basis for engineering optical nonlinearity.

In our experiments, we encapsulate the homotrilayer of WSe$_2$, exfoliated from bulk 2H WSe$_2$ crystals, inside two layers of hBN (~15 nm thick) and investigate their optical response. In a dual-gate geometry, we use gate voltages applied to the top and bottom graphite to individually control the overall doping levels in the trilayer and the displacement field across it (**Fig. 1a & b**). **Figure 1c** shows the photoluminescence spectra of the sample by varying the electric field while keeping the samples undoped. We observe strong emission from excitons X$_I$, whose energy linearly shifts with electric field from 1.58 eV to 1.5 eV, and PL peaks at 1.71 eV, 1.52 eV, 1.5 eV that remains constant with varying electric field. We attribute the peak at 1.71 eV to intralayer momentum direct exciton X$_A$, corresponding to the K-K transition, based on their strong absorption which will be discussed later. The lower energy of X$_I$ indicates that they are momentum-indirect excitons at the band edge, corresponding to transition across the smaller indirect gap between Γ and Q valleys[33-35]. From the slope of the linear Stark shift, we estimate the electric dipole moment of X$_I$ to be around $0.65\ nm \cdot e$. The corresponding vertical displacement of the electron-hole pair in X$_I$ is roughly half of the value of the distance between the top and bottom tungsten layers, indicating the electrons or holes are partially layer delocalized. The PL peaks at the energy of 1.55 eV and 1.52 eV could originate from momentum-indirect or defect-bound excitons with no dipole moment, which becomes dark under a high electric field when X$_I$ energy becomes smaller than 1.55 eV.

Next, we measure the reflectance of the trilayer under an electric field (**Fig. 1d**). In addition to the intralayer X$_A$ excitons at 1.71 eV, we observe an additional strong reflectance contrast at 1.78 eV

(IX$_D$), which exhibits a substantial Stark effect of almost 100 meV. The finite reflection contrast and linear Stark effect of IX$_I$ suggest that it corresponds to interlayer exciton at the direct K-K transition with larger oscillator strength than those momentum-indirect excitons observed in PL. From the slope of the Stark effect, we estimate the electron-hole displacement to be 1.35 $nm$. Interestingly, as the energy of IX$_D$ approaches that of the intralayer exciton X$_A$ under a higher electric field, we observe an apparent anti-crossing behavior of X$_A$ and IX$_D$ near the electric field of 0.05 V/nm (see data from an additional device in **Fig. S1**). We note that the levels are not fully avoided and there is always finite reflection from X$_A$ at 1.71 eV for all electric fields. This suggests that only certain part or species of the X$_A$ which sit at both the top and bottom layer interacts with IX$_D$, which we will discuss later. To quantitatively understand the avoided crossing, we extract the exciton energies by fitting reflectance spectra and then model the anti-crossing using a simple coupled oscillator model. From such a model, we estimate a coupling strength of W = 10 $\pm$ 2 meV between X$_A$ and IX$_D$ (**Fig. S2**).

We further investigate how electrostatic doping influences the optical response of the excitons. **Fig. 1e** shows the reflectance spectra of trilayer WSe$_2$ as we vary the doping levels while keeping the electric field at zero. With doping, the reflectance from neutral interlayer excitons diminishes as they lose their oscillator strength, while the charged intralayer excitons emerge, i.e. trions or Fermi polarons, which redshifts with increasing doping levels. Because we focus on the highly doped regime where excitons interact with a large number of carriers, we refer these charged excitons to Fermi polarons in later discussions[12]. In PL, the emission of the charged excitons also redshifts with doping but becomes more intense than the intrinsic X$_A$ (**Fig. S3**). Interestingly, as we apply an electric field across the doped sample, the charged interlayer excitons can acquire oscillator strength when they are in resonance with the intralayer excitons, and this leads to clear avoided crossings at different doping levels (see **Fig. S1c, d**). Following a similar analysis for the undoped case, we find that the coupling strength remains close to W ~10 meV on both the hole and electron doping sides.

Next, we study the excitons' nonlinear optical response by measuring the trilayer's reflectance spectra under different laser pumping. **Fig. 2a-c** shows the reflectance spectra of the sample, probed with a broadband halogen lamp, while we excite the system with a 635nm continuous wave (CW) laser of different power. When the trilayers are electron-doped or intrinsic, optical pumping does not alter the reflectance spectra significantly. Intriguingly, however, in the hole-doped regime, optical pumping leads to a dramatic blueshift of the exciton energies, on the order of a few millielectronvolts under microwatts excitation (**Fig. 2b**). We note that trilayer PL signals are more than four orders of magnitude weaker than the reflected light and are negligible in the measured spectra. **Fig. 2d** shows the reflectance spectra change induced by the optical pumping, relative to the reflectance spectra without pumping, under symmetric gating where we vary the doping concentrations without applying an electric field. We observe a striking asymmetry between the electron and hole sides. Furthermore, we can also probe the optical nonlinearity by resonantly exciting $X_A$ with a pulsed laser (in the wavelength range of 718 to 730 nm with ~100 ps pulse duration with 40 MHz repetition rate) and measure the spectrum of the reflected pulse. Under this pulsed excitation scheme, we observe qualitatively similar nonlinear behaviors where the blueshift is only observed on hole-doped side (**Fig. S4**). **Figure 2e** shows how the reflectance contrast of the $X_A^+$ evolves as we increase the average power of the resonant excitation laser. A blueshift of as much as 10 meV is induced under tens of microwatts of optical pumping. Notably, the pulsed and

CW lasers produce a similar order of magnitude of peak shift under the same average power despite the orders of magnitude difference in the peak powers (**Fig. S5**).

To further understand our observations, we investigate how an applied electric field influences optical nonlinearity. First, we measure the reflectance of the $X_A^+$ and $X_A^-$ as we change the electric field but fix the doping levels (**Fig. 3a & b**). With an increasing electric field, the energy of the $X_A^+$ blueshift quadratically at a smaller electric field and then switches to redshift above an electric field of ~0.05 V/nm (**Fig. 3a**). In contrast, the energy shift of the $X_A^-$ is much smaller. Interestingly, the electric field also dramatically changes the nonlinear behaviors of the excitons. **Figure 3c** shows the relative change of the reflectance spectra by optical pumping with a fixed hole doping density. At an electric field around 0.05 V/nm, we observe a clear flip in the color contrast map, which corresponds to a switchover from an optically induced blueshift to a redshift of $X_A^+$. We note that the amount of blueshift and redshift are similar, around 10 meV (**Fig. 3d**).

The observed large Stark effect and anti-crossing can be understood by examining the crystal and band structure of trilayer $WSe_2$. In trilayers, each monolayer is rotated by 180 degrees with respect to the neighboring layer, resulting in alternating K and K' points between layers[36] (**Fig. 4a, b**). Examining the Wannier wavefunction obtained from density functional calculations suggests that the strength of electron tunneling is much weaker than that of hole tunneling. The sizeable spin-orbit coupling in the valence band dictates that the direct tunneling between the neighboring layers would be much weaker than that between the top and bottom layers across the middle layer (**Fig. 4a, b**). Such tunneling between the top and bottom layer leads to finite oscillator strength of interlayer K-K excitons $IX_D$ and their avoided crossing with intralayer excitons $X_A$[37]. The coupling strength between $X_A$ and $IX_D$ extracted from our experiments is consistent with the interlayer coupling strength of holes at K valley[35]. This picture is further corroborated by our measured dipole moment of $IX_D$ being close to the distance between the top and bottom layers. It can be also seen that $IX_D$ does not couple to intralayer excitons in the middle layer, explaining our observation that the level crossing at $X_A$ is not fully avoided.

To understand the optical nonlinearity in the hole-doped $WSe_2$ trilayers, we examine how intralayer excitons $X_A$ interact with other elementary excitations in the system, such as $IX_I$ and free carriers[12,38,39]. First, the large blueshift of intralayer excitons cannot be simply explained by heating or carrier injection from the laser excitation, since both effects result in redshift of the intralayer excitons (**Fig. 1**). On the other hand, the interactions among purely intralayer excitons are repulsive but weak, which scales linearly with density with a coefficient of $\sim E_B R^2 \sim 5\times 10^{-12}$ meVcm$^2$, where $E_B \approx 500$ meV is the exciton binding energy and $R$ (~1nm) denotes the exciton Bohr radius[40,41]. Although the hybridization between intra- and inter-layer excitons could enhance the nonlinearity of $X_A$ by introducing dipolar interactions, the experimental excitation power is unlikely to produce sufficient exciton density to explain the experiments given the short lifetime of intralayer excitons (~picoseconds)[42,43]. In fact, our observation that a CW and pulsed laser produces a similar order of magnitude of blueshift under equal average power suggests that the interaction involves processes happening on a nanosecond timescale that is much longer than the intralayer exciton lifetime. Therefore, the observed nonlinearity likely originates from the interactions of intralayer excitons with species of much higher population, such as momentum-indirect excitons ($X_I$) and free carriers (**Fig. 4c**).

While there could be relatively strong interactions between $X_A$ and $X_I$ due to their acquisition of electric dipoles, $X_A$-$X_I$ interaction alone is unlikely to explain the observation since the nonlinearity is absent in the intrinsic case. Instead, the nonlinearity is possibly a result of the strong interactions between intralayer excitons and free carriers. In particular, excitons created by optical pumping may induce a valley population imbalance of resident carriers between K and Γ valleys, via mechanisms such as exciton-carriers scattering (**Fig. 4d**). Importantly, under zero electric field, the energy difference between K and Γ is rather small in trilayers, and Γ point is spin-degenerate. As a result, holes at the Γ point could be efficiently scattered into the K valley by excitons such as $X_A$ and $X_I$, via Coulombic and exchange interactions[44,45]. This creates a net accumulation of valley population at K (and K'), which induces phase space filling and, consequently the observed blueshift. Such a blueshift induced by exciton-charge interactions does not occur in the intrinsic regime because of the lack of free carriers. Meanwhile, free electrons will likely experience a much higher energetic barrier for population transfer into K valleys due to the much larger energy splitting of Q and K valleys in the conduction band. This explains the observed strong asymmetry of optical nonlinearity on the electron vs. hole side. Another possible mechanism for optically induced valley polarization could be that the accumulation of $X_I$ excitons with finite dipole moment creates an effective displacement field across the trilayer by introducing a relative energy shift between K and Γ. In both proposed mechanisms, $X_I$ can play an important role in the exciton-carrier interaction, which could underly the dynamics of the nonlinearity (**Figs. 2d and S5**). Based on the lifetime of $X_I^+$ and the sample's absorption at the laser excitation, we estimate the density of the $X_I^+$ to be at least ten times smaller than the resident carriers, but comparable to the required valley polarization to induce the observed blueshift by phase space filling.

The above picture is also consistent with the optically induced redshift under large electric field on the hole side (**Fig. 4e, f**). In particular, the applied electric field changes the valence band edge from $\Gamma$ to K[35]. Under strong optical excitation, the same processes, such as exciton-hole scattering or effective displacement field created by $X_I$ excitons, can induce a population transfer of holes from K to $\Gamma$, which reduces the energies of the Fermi polarons (**Fig. 4f**). While these constitute a consistent picture for explaining the experiments, elucidating the rather complex interactions between intralayer, interlayer excitons and charges requires further theoretical studies and will be of significant future interest.

It is noteworthy that in monolayers, optical pumping with circular polarized light can polarize carriers in K vs. -K valleys[35, 42–48]. The generation of valley imbalance in these cases has been attributed to mechanisms, such as different inter-valley vs. intra-valley carrier relaxation rates, and different indirect excitons and spin-forbidden dark excitons relaxation rates[49,50]. A notable difference in the trilayer case is that the lack of spin-orbit coupling at $\Gamma$ facilitates more straightforward intervalley scattering. Lastly, we also briefly discuss the difference of our observations from recent experiments on bilayer $MoS_2$[30,51,52]. In bilayers, interlayer excitons acquire significant oscillator strengths by hybridization with the intralayer B but not A excitons ($X_A$), and previous reports focused on the nonlinearities of the interlayer excitons. Remarkably, our observed nonlinearity of intralayer excitons is more than ten times stronger than that of dipolar interlayer excitons in bilayers[51,52].

Our results demonstrating highly nonlinear excitons with large oscillator strengths open avenues for engineering the interactions between excitons and carriers in atomically thin semiconductors and their heterostructures to explore strongly interacting many-body physics and develop new applications. By designing the atomic and electronic structures of the heterostructures, one may engineer the strong interactions between bright, dark excitons, and free charges[53,54]. The excitonic and free charge populations are highly tunable, enabling the exploration of many-body physics in a hybrid Fermi-Bose system[17–19]. These strongly interacting optical excitations can be used to realize active nonlinear metasurface based on spatial confinement of excitons by moiré superlattice and local electrostatic gate[15,55-57]. Combining strong nonlinearity with spatial confinement could also pave the way for exploring quantum optical effects, including nonclassical light sources and few-photon nonlinearity[16,58]. Finally, the demonstrated optical control of exciton resonances could enable novel nonlinear optoelectronic devices such as all-optical switching, nonlinear optomechanical resonators[41,59], and optical limiting devices[3, 60].

**Acknowledgements:**

This research is supported by the U.S. Department of Energy, Office of Science, Office of Basic Energy Sciences Early Career Research Program under Award No. DE-SC-0022885 and the National Science Foundation CAREER Award under Award No. DMR-2145712. This research used Quantum Material Press (QPress) of the Center for Functional Nanomaterials (CFN), which is a U.S. Department of Energy Office of Science User Facility, at Brookhaven National Laboratory under Contract No. DE-SC0012704.

**Author contributions**

Y.Z. and L.G. conceived the project. L.G. fabricated the samples and performed the experiments. L.Z., R.N., S.P., and H.J. assisted with sample fabrication. L.Z. and R.N. helped with optical measurements. M.X., D.S.W. M.H. and Y.Z. contributed to the data analysis and theoretical understanding. T.T. and K.W. provided hexagonal boron nitride samples. L.G. and Y.Z. wrote the manuscript with extensive input from the other authors.

**Competing financial interest**

The authors declare no competing financial interests.

**Additional Information**

Supplementary information is available in the online version of the paper. Reprints and permission information is available online at www.nature.com/reprints. Correspondence and requests for materials should be addressed to youzhou@umd.edu.

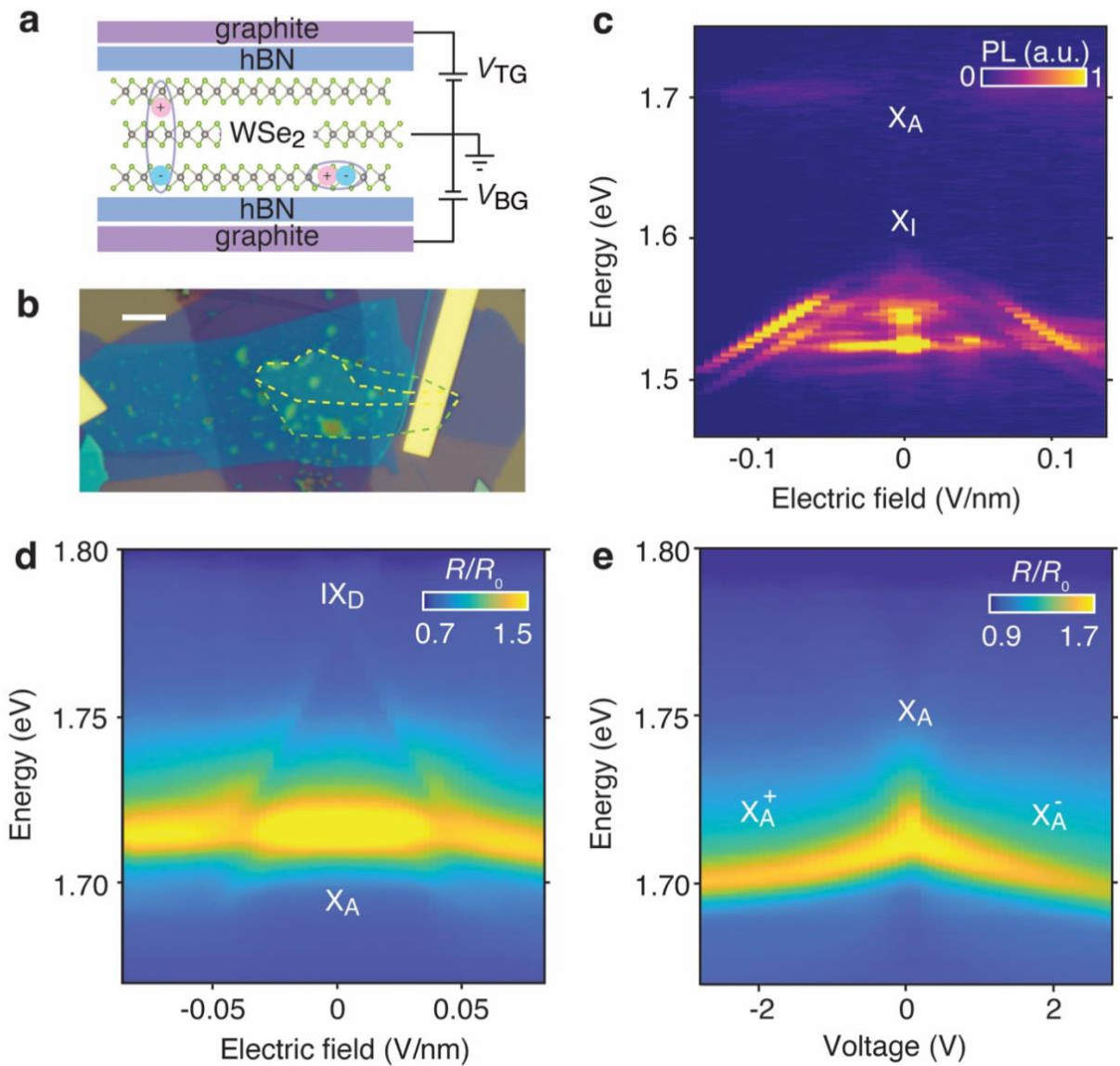

**Figure 1. Dual-gated WSe$_2$ homotrilayer van der Waals heterostructure and their optical characteristics under gating at $T$ = 4 K. a,** a schematic of the trilayer vdWs heterostructure. The homotrilayer WSe$_2$ is encapsulated with two hBN of 15~20 nm thick. **b,** an optical image of the homotrilayer WSe$_2$ device (Scale bar: 5 $\mu m$). The trilayer region is indicated by the yellow dashed line. The green dashed line corresponds to the neighboring homobilayer region. **c,** Photoluminescence map of the WSe$_2$ trilayer with electric field. The bright emission with a stark shift from 1.5 to 1.58eV with electric field corresponds to the indirect exciton X$_I$. The upper weaker emission at 1.7eV corresponds to the momentum direct K-K intralayer exciton X$_A$. **d,** Reflectance spectra with electric field of the WSe$_2$ trilayer. The high energy momentum direct K-K interlayer exciton shows a stark shift of ~100 meV and coupled with X$_A$ exciton when they become energy degenerate around the electric field of 0.05 V/nm. **e,** Doping density dependent reflectance spectra of the WSe$_2$ trilayer. The negative voltage corresponds to the hole doped side. With increasing doping concentration, the intralayer trion or Fermi polaron (X$_A^-$/X$_A^+$) shifts toward lower energy.

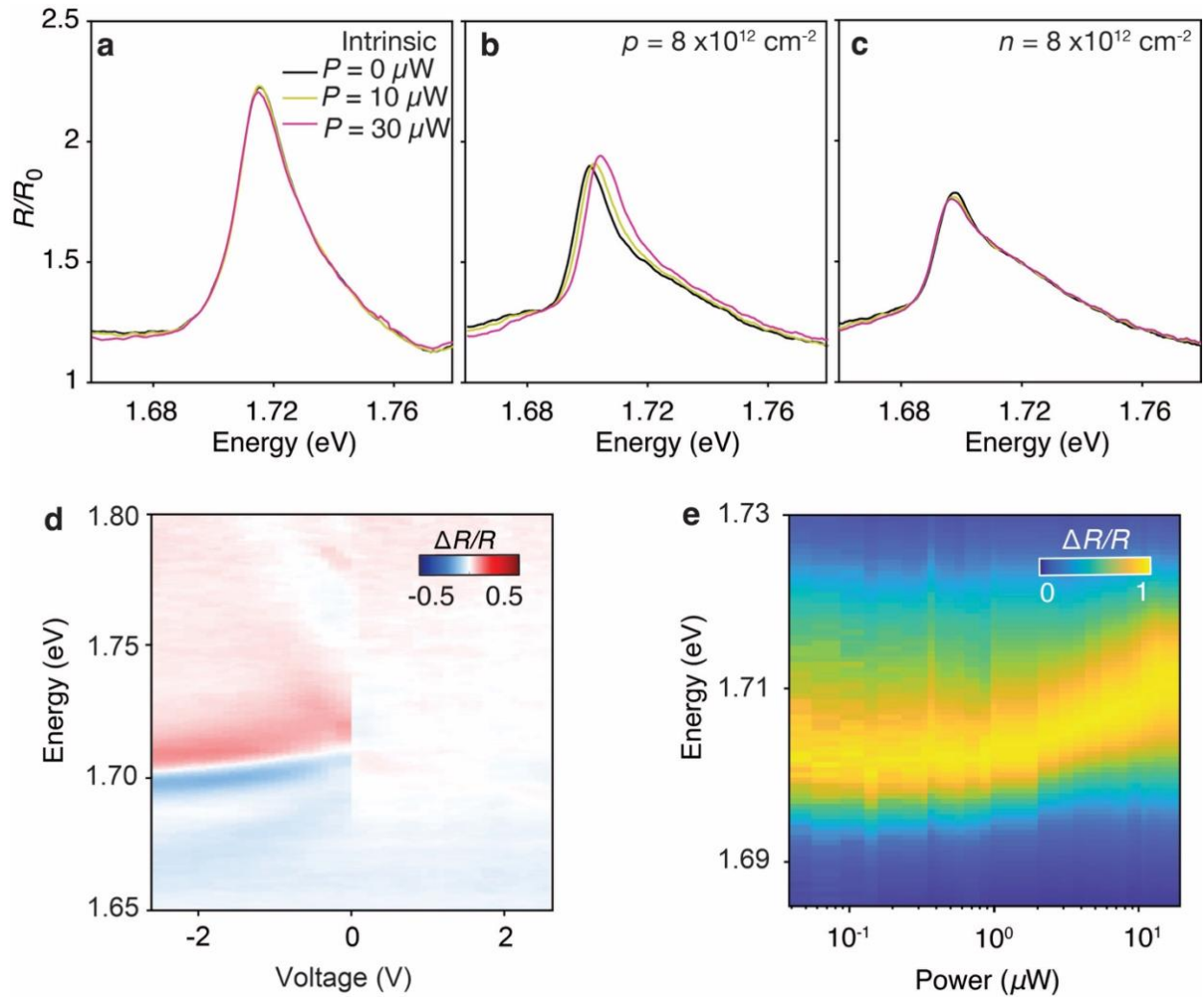

**Figure 2. Nonlinearity in hole doped homotrilayer WSe$_2$ at $T$ = 4 K . a, b, c,** Reflectance contrast $R/R_0$ of the trilayer under 0 $\mu$W, 10 $\mu$W, 30 $\mu$W CW (635nm) laser excitation with different doping levels, where $R_0$ is the reflectance of a reference region near the trilayer region with bare hBN on SiO$_2$ on the sample. Doping densities of hole and electron are kept at 5.6× $10^{12} cm^{-2}$ in both **a** and **b**. Intriguingly the exciton blueshifts when the sample is hole doped. **d,** Relative change in the reflectance induced by 30 $\mu$W of CW laser pumping under different doping. The color map is obtained by normalizing the reflectance change induced by the CW excitation with respect to the reflectance without optical pumping, $\Delta R/R = \frac{R_{(30\mu W)}}{R_{(No\ Pump)}} - 1$. The color contrast (red on top, blue on bottom) refers to the peak blue shift. The negative voltage corresponds to the hole doped region. **e,** Peak shift of the $X_A^+$ as a function of the pulsed laser excitation (718-720nm, resonant with $X_A^+$) power. The hole doping density in the system is $8 \times 10^{12} cm^{-2}$.

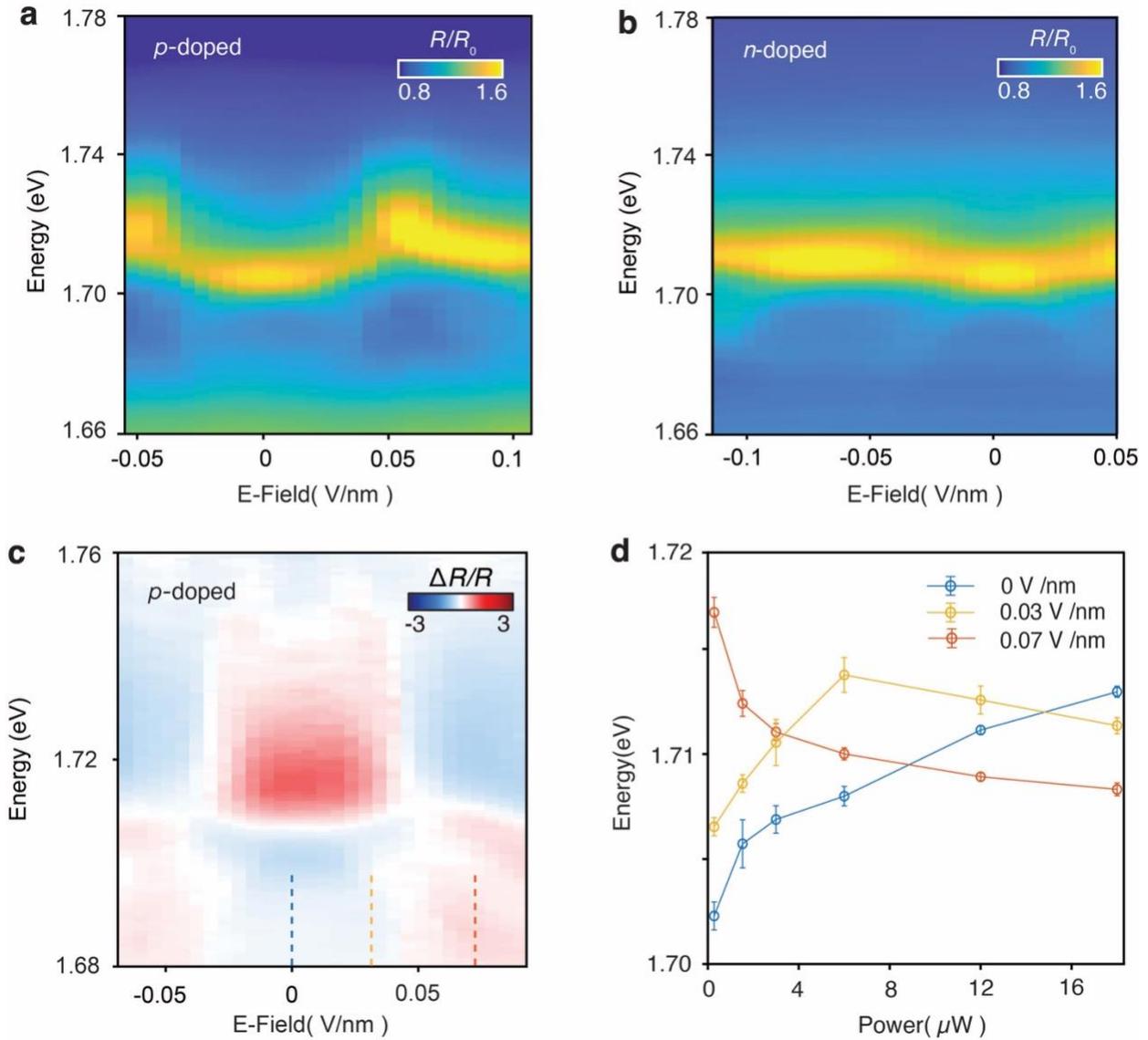

**Figure 3. Electric-field dependent exciton energy and nonlinearity in homotrilayer WSe$_2$ at T = 4 K. a, b,** Electric field dependence of the intralayer Fermi polaron reflectance contrast $R/R_0$ in trilayer with a (**a**) hole and (**b**) electron doping density of $5.6\times 10^{12} cm^{-2}$. **c,** Reflectance change induced by a pulsed laser excitation of 12 $\mu$W power. The color map is obtained the same way as Fig.2d. Under a small electric field, $X_A^+$ shows a blueshift but it begins to redshift under excitation at higher electric field. **d,** Line plots of the $X_A^+$ power dependent peak shift under different electric field. The $X_A^+$ shows a blue shift in a regime of 10 meV under zero applied electric field and flip to redshift which shows compatible amount as the blueshift. The corresponding electric fields for these linecuts are indicated by the dashed lines in (**c**).

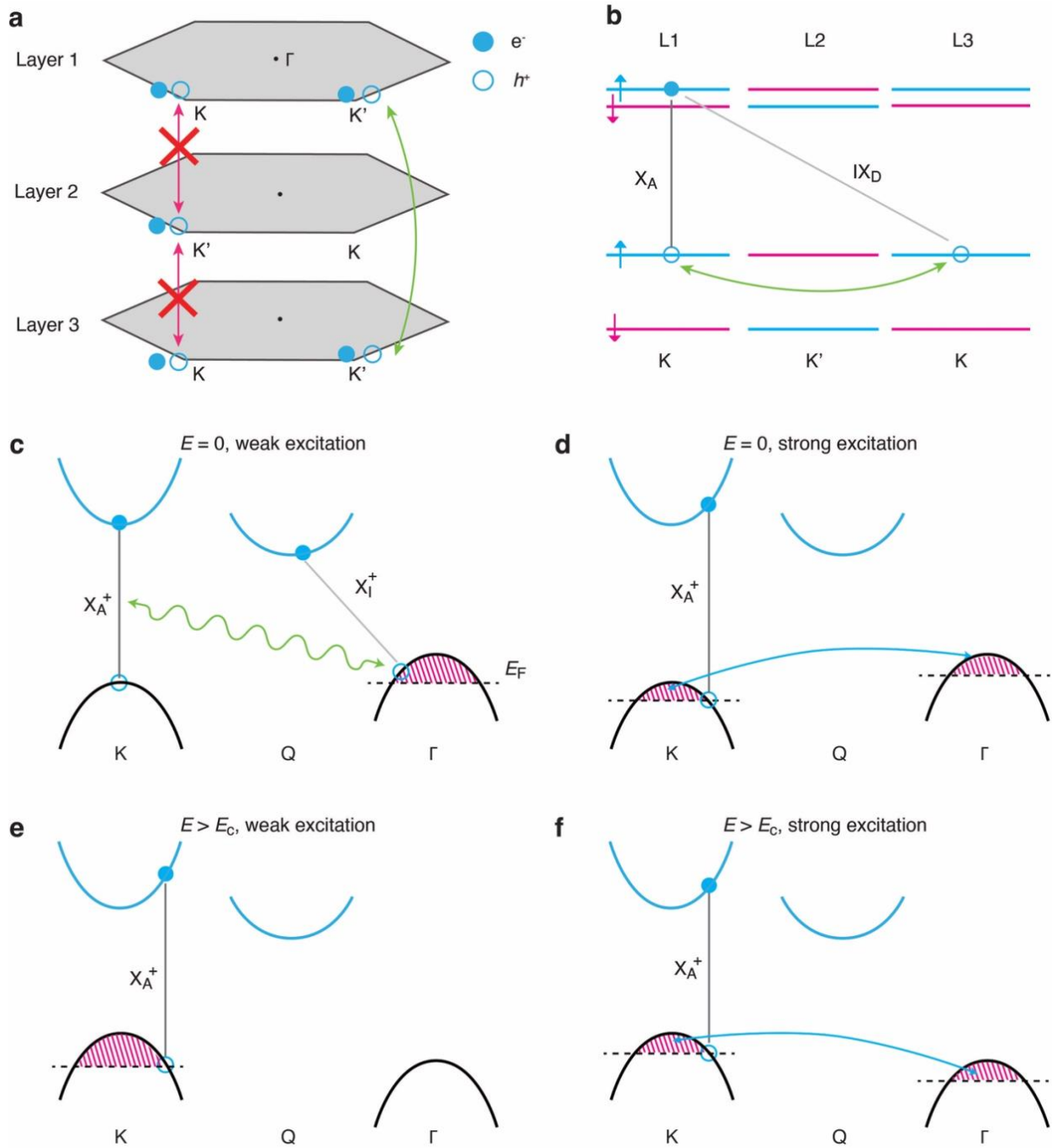

**Figure 4. Electronic band structure of trilayer WSe$_2$. a,** The crystal structures of natural trilayer WSe$_2$ dictates alternating K and K' valleys among neighboring layers. The strong spin-orbit coupling of holes leads to weak tunneling among neighboring layers but strong tunnelling

between the top and bottom layer. **b,** The tunnelling of holes between top and bottom layer results in the hybridization of intralayer K-K excitons $X_A$ and interlayer K-K excitons $IX_D$. **c, d,** Band structures and carrier populations of hole-doped trilayer $WSe_2$ in the absence of electric field. **c,** Optical excitation generates both momentum intralayer $X_A^+$ and momentum indirect $X_I$ of higher population. Intralayer Fermi polaron $X_A^+$ can interact with $X_I$ and the free holes in the system. **d,** Under strong optical excitation, the interaction between intralayer excitons and free charges can induce a population transfer of carriers from the $\Gamma$ to the K valley. The energy difference between $\Gamma$ and K is small. The additional free carriers in K valley leads to phase space filling and optically induced blueshift of $X_A^+$. **e, f,** An electric field induces a shift of the valence band edge from $\Gamma$ to K. Under strong optical pumping, a population imbalance is induced with increased carriers at the $\Gamma$ point, which leads to the optically induced redshift of $X_A^+$ at higher electric field.

## Methods:

### Device fabrication

Graphite, hBN flakes are mechanically exfoliated from the bulk crystals onto the silicon chip with SiO$_2$ layer. Some of the exfoliated homotrilayer WSe$_2$ flakes were provided by the Quantum material press (QPress) facility in Center for functional nanomaterials (CFN) at Brookhaven national laboratory (BNL). The thickness of the hBN flakes and WSe$_2$ layer numbers are estimated based on the colour contrast under the optical microscopy. The heterostructure is assembled in a transfer station built by Everbeing Int'l Corp., which use PDMS(Polydimethylsiloxane) and PC (Polycarbonate) as stamp and transfer all the flakes in a dry transfer method onto a silicon chip with 285nm SiO$_2$ layer. Then the electrical contacts are patterned by electron-beam lithography and a liftoff process where we deposited 5nm of Cr and 80nm of Au by thermal evaporation.

### Optical spectroscopy

The optical measurements were performed in our home-built confocal microscope with Attodry 4K cryostat. The apochromatic objective equipped in the chamber has numerical aperture NA=0.82. The PL measurement is performed with a 635nm diode laser excitation. The reflectance measurement is performed using either a halogen lamp (from Thorlabs) or a supercontinuum white laser (from YSL Photonics Inc.) as the excitation source. The diode laser has a diffraction-limited spot size while the beam diameter of the white laser is slightly larger, around $1 \mu m$. The white laser has a pulse duration of ~60 ps with variable repetition rate up to 40 MHz. Our power-dependent reflectance is measured under both continuous wave and pulsed excitation. In CW measurements, we illuminate the sample with the halogen lamp as the probe and use the CW (635nm) diode laser as excitation with a power ranging from 0.02 to 30 microwatts. In the pulsed resonant excitation case, we excite the system with a supercontinuum white laser filtered to 718-730nm wavelength range. We vary the incident power and directly measure the reflected white laser signal from the sample. In both cases, our reflectance spectral is normalized by dividing the reflected light intensity from the sample trilayer area by the reflected light intensity from the nearby bare hBN on SiO$_2$ area. The spectra is measured by a Horiba iHR320 spectrometer using a 300mm/line grating and a Synapse-Plus back-illuminated deep depletion CCD camera.

**Doping Density and Electric field**

The doping density and electric field are determined by considering the heterostructure as a parallel capacitor[30]. The applied electric field is calculated as $E = \frac{D}{\varepsilon_0 \varepsilon_{WSe2}}$, while the D displacement field is determined by $D = \frac{1}{2}(C_{Top} \cdot \Delta V_T - C_{Bottom} \cdot \Delta V_B)$. The top and bottom capacitance are given by $C_{Top(Bottom)} = \frac{\varepsilon_0 \varepsilon_{hBN}}{t_{hBN}}$. $\Delta V_T$ and $\Delta V_B$ is the applied top and bottom gate voltage relative to the offset voltage to the band edge, respectively. The total doping density in the system can be determined as $n = \frac{1}{e} \cdot (C_{Top} \cdot \Delta V_T + C_{Bottom} \cdot \Delta V_B)$. We use $\varepsilon_{WSe2} = 7, \varepsilon_{hBN} = 3$ in our case. The thicknesses of hBN layers are extracted by atomic force microscope measurements.

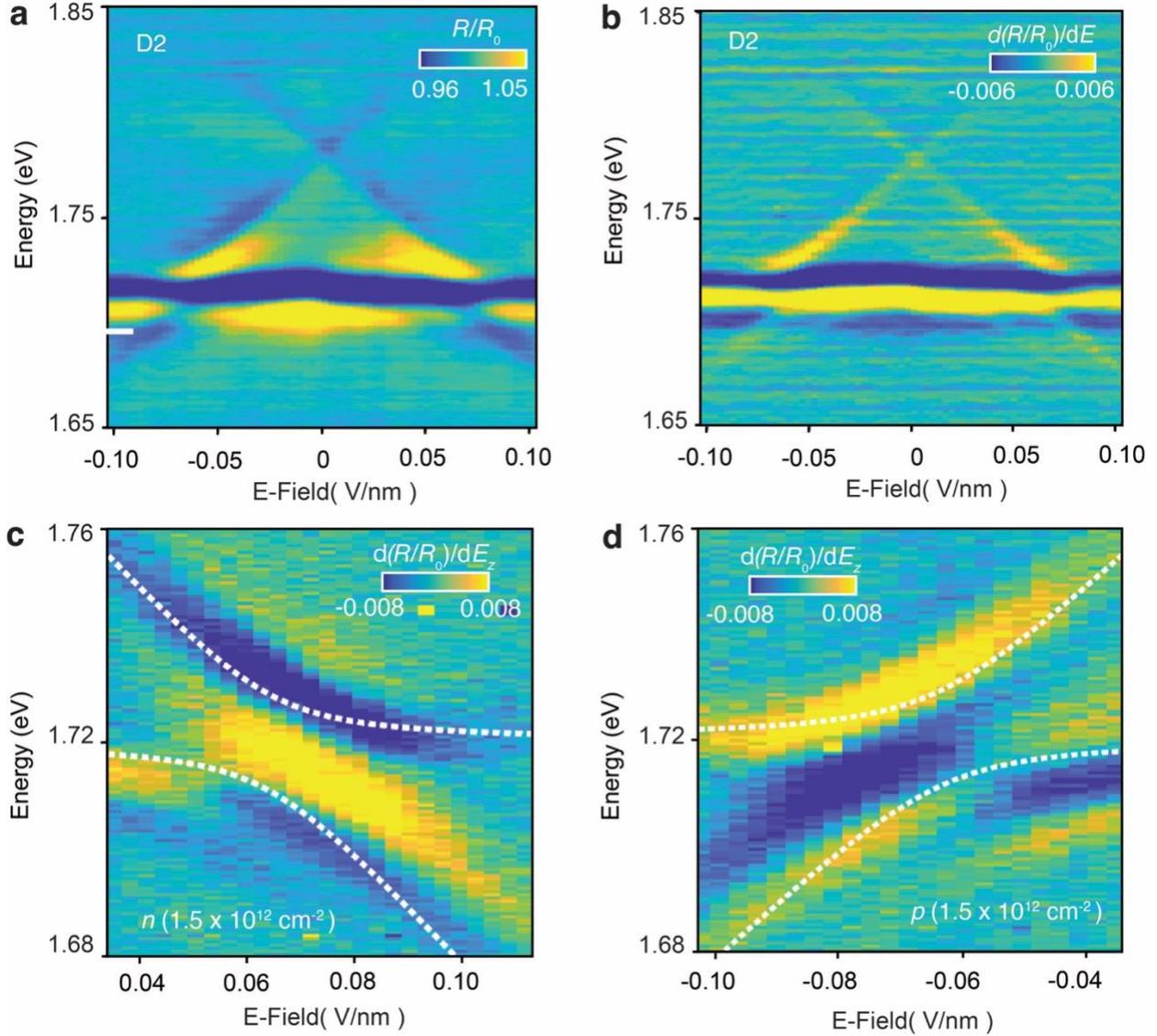

**Figure S1. Anti-crossing between IX$_D$ and X$_A$ for various doping concentrations in device D2. a,** Reflectance spectra ($R/R_0$) as a function of the electric field in the intrinsic regime. **b,** Differential reflectance spectrum (d($R/R_0$)/dE) as a function of the electric field. The anti-crossing takes place at ~0.05 V/nm, consistent with device D1. **c, d,** Zoom-in view of the differential reflectance under electron (**c**) and hole(**d**) doped with an applied electric field. The white dashed lines represent the energies fitted with a two-level model. The fitted coupling strength **W** for electron and hole-doped side is around ~10 meV and comparable with that in intrinsic trilayer.

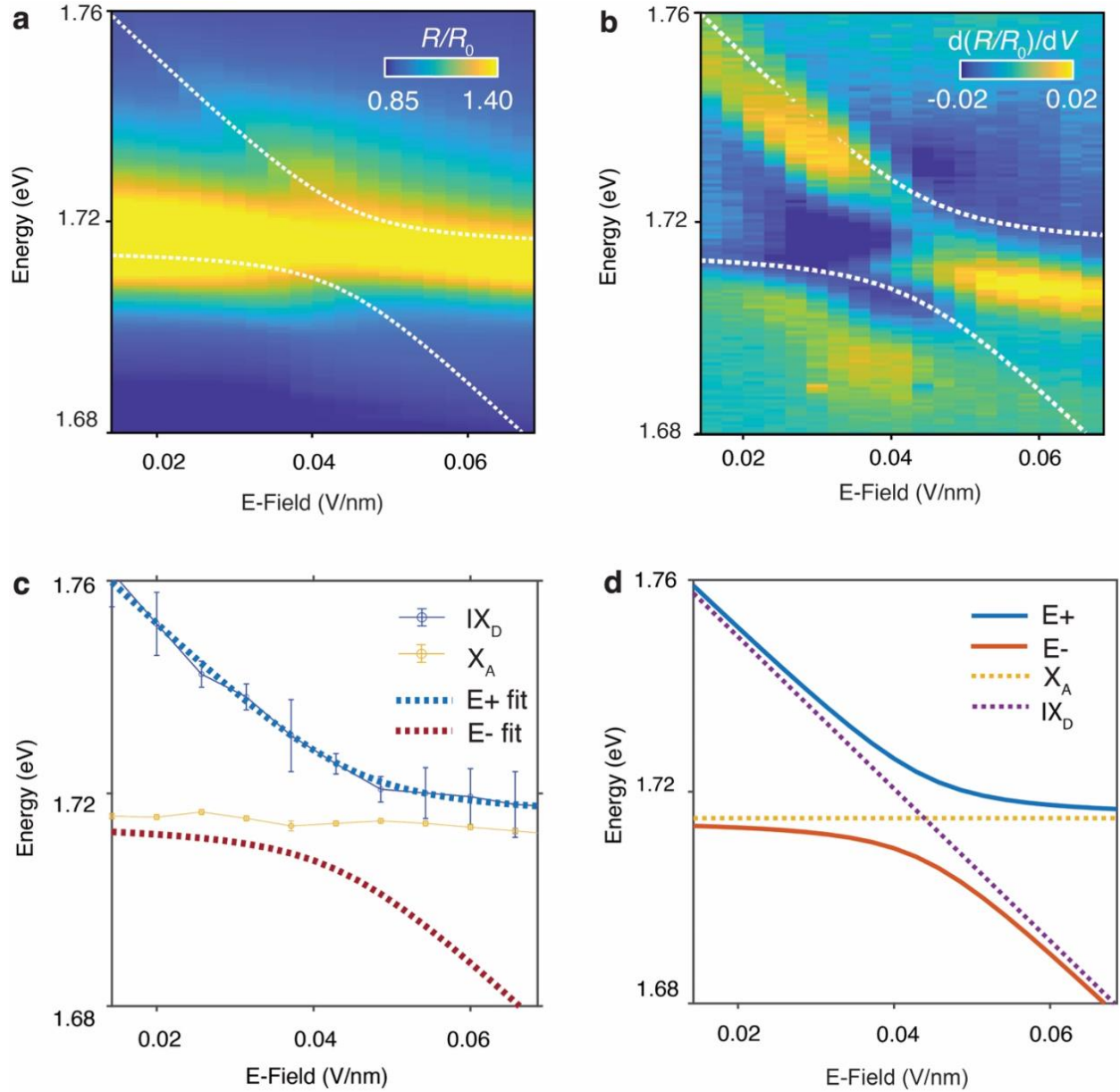

**Figure S2. Analysis of anti-crossing between IX$_D$ and X$_A$ in device D1. a,** Reflectance spectra ($R/R_0$) as a function of the electric field near the anti-crossing region. **b,** Voltage derivative of reflectance spectra, d($R/R_0$)/d$V$. The white dashed lines in (a) and (b) represent the energies fitted with a two-level model. **c, d,** We study the anti-crossing between the IX$_D$ and X$_A$ based on a two-level system with a Hamiltonian:

$$H = \begin{pmatrix} E1 & W \\ W & E2 \end{pmatrix}$$

where $E_1$ and $E_2$ are the unperturbed energies of the IX$_D$ and X$_A$, respectively, and $W$ is the coupling strength. The new eigenvalues can be expressed as:

$$E_\pm = \frac{1}{2}(E_1 + E_2) \pm \frac{1}{2}\sqrt{(E_1 - E_2)^2 + 4|W|^2}$$

where $E_\pm$ correspond to the energies of the two branches. In (**c**), we extract the peak positions $E_\pm$ by fitting the reflectance spectra with the Lorentzian function. We then set $E_2$ to be 1.715eV, which is the mean energy of X$_A$, and keep it as a constant. $E_1$ is calculated based on the IX$_D$ energy at zero electric fields and the stark shift. The stark shift slope $k$ is estimated to be -1.436 eV/V in this particular device. The fitted anti-crossing is shown in (**d**) with a fitting parameter of $W = 10 \pm 2\ meV$.

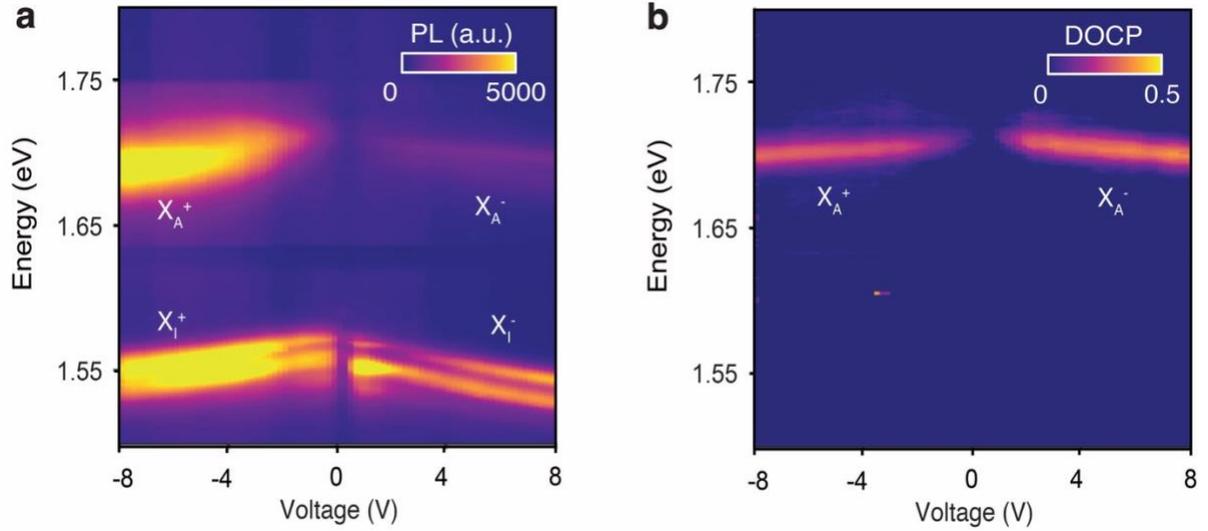

**Figure S3. a, Doping-dependent photoluminescence of the trilayer WSe$_2$ at $E_z = 0$ taken from D2 at 4 K. b, Degree of circular polarization (DOCP) of the $X_A^+$ and $X_A^-$.** The bright emission in the range of 1.5~ 1.6 eV corresponds to the momentum indirect trion/Fermi polaron. In contrast, the higher energy emission around 1.7eV corresponds to the momentum direct (K-K) intralayer trion/Fermi polaron. Both charged excitons $X_I$ and $X_A$ exhibit a redshift with increasing doping density.

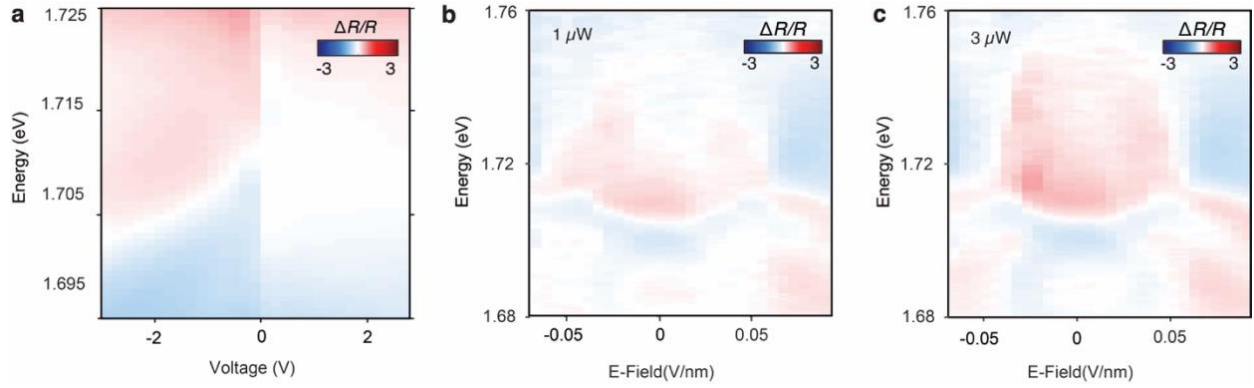

**Figure S4. a,** Relative change in the reflectance induced by 1 $\mu$W of resonant at resonant (718 to 730 nm) pulsed laser excitation under different doping. The color map is obtained by normalizing the reflectance change induced by the resonant excitation with respect to the reflectance without optical pumping, $\Delta R/R = \frac{R_{(1\,\mu W)}}{R_{(0.1\,\mu W)}} - 1$. The pulse has ~100 ps duration with 40 MHz repetition rate. **b, c,** Reflectance change induced by a pulsed laser excitation power of 1 $\mu$W(**b**) and 3 $\mu$W(**c**), as a function of electric field, under hole doping. Under a small electric field, $X_A^+$ shows a blueshift, but it begins to redshift under excitation at a higher electric field. With increasing power, this transition point shifts to a lower electric field.

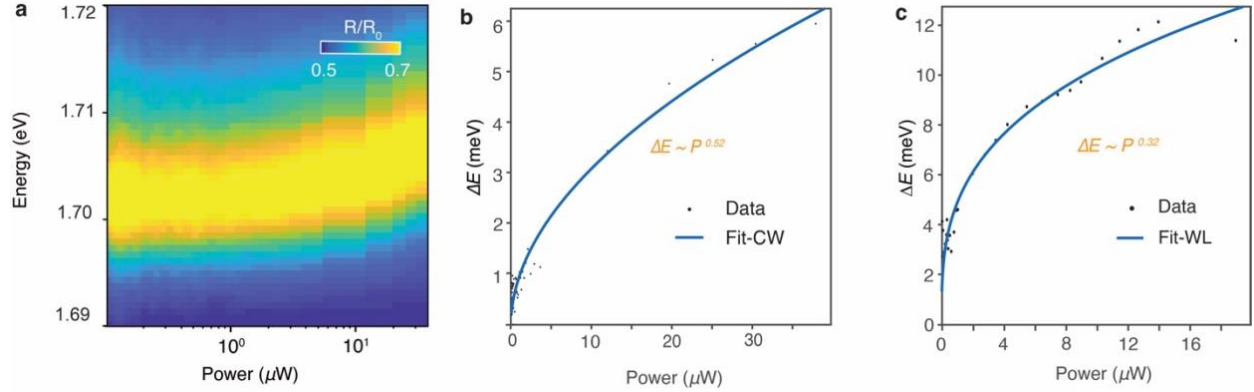

**Figure S5. Power-dependent blueshift of $X_A^+$ under CW laser (a,b). and pulsed white laser excitation(c).** In **b, c,** we fit the blue shift amount as $\Delta E = a \cdot P^{(b)}$. The fitting for CW laser and pulsed laser yields a coefficient of $b$ as 0.52 with an R-square of 0.9516 for the CW laser and $b$ of 0.32 with an R-square of 0.9317 for the pulsed laser, which shows a sublinear response for $X_A^+$ as power. The hole doping density is kept at 8 x $10^{12}$ cm$^{-2}$. The peak blue shift is around 6 meV with a maximum excitation power of 40 $\mu w$.

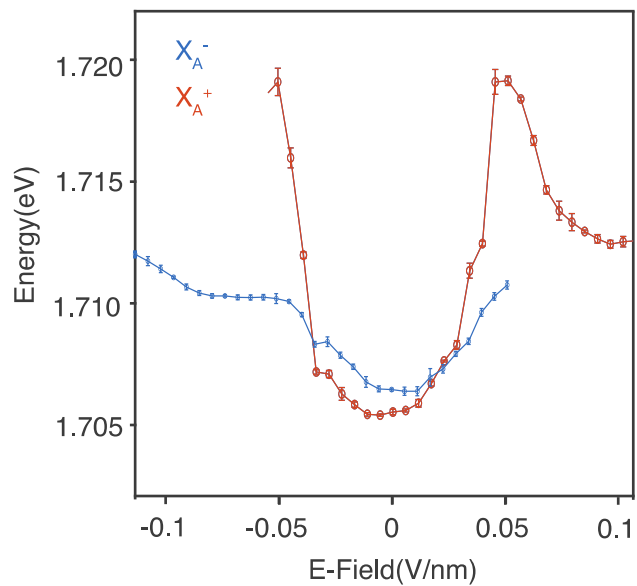

**Figure S6. Energy shift of $X_A^-$ and $X_A^+$ with applied electric field with constant doping.** The electron and hole doping density are both kept at $5.6\times 10^{12} cm^{-2}$. The peak position is obtained by fitting the reflectance spectral with a Lorentzian model.